\documentstyle[twocolumn,aps,psfig]{revtex}
\begin{document}
\draft
\twocolumn[\hsize\textwidth\columnwidth\hsize\csname @twocolumnfalse\endcsname
%
%
%

\title{Phase Separation in Electronic Models for
Manganites}

\author{ S. Yunoki$^1$, J. Hu$^1$, A. L. Malvezzi$^1$, A. Moreo$^1$, 
N. Furukawa$^2$, and E. Dagotto$^1$}

\address{$^1$ National High Magnetic Field Lab and Department of Physics,
Florida State University, Tallahassee, FL 32306}
\address{$^2$ Institute for Solid State Physics, University of Tokyo,
Roppongi 7-22-1, Minato-ku, Tokyo 106, Japan}

\date{\today}
\maketitle

\begin{abstract}
The Kondo lattice Hamiltonian with ferromagnetic Hund's coupling as a
model for manganites is investigated.
The classical limit for the spin of the (localized)  $t_{2g}$ electrons is
analyzed on lattices of dimension 1,2,3 and $\infty$ using
several numerical methods.
The phase diagram at low temperature is presented. A regime is
identified where phase separation occurs between hole
undoped antiferromagnetic and hole-rich ferromagnetic regions. 
Experimental consequences of this novel regime are discussed.
Regions of incommensurate
spin correlations have also been found. 
Estimations of the critical temperature in 3D  are compatible
with experiments.

\end{abstract}
\vskip2pc]
\narrowtext

%
%

The phenomenon of
colossal magnetoresistance in  metallic oxides such as
${\rm La_{1-x} Sr_x Mn O_3}$ 
has recently attracted considerable
attention~\cite{jin} due to its potential technological applications.
A variety of experiments have revealed that  oxide
manganites have a rich phase diagram~\cite{phase} with regions of antiferromagnetic
(AF) and ferromagnetic (FM) order, as well as charge ordering, and
a peculiar insulating state above the FM critical temperature, $T_c^{FM}$.
Recently, $layered$ manganite compounds 
${\rm La_{1.2} Sr_{1.8} Mn_2 O_7}$
have also been synthesized~\cite{moritomo} with properties similar to
those of their 3D counterparts.

The appearance of ferromagnetism at low temperatures can be explained using
the Double Exchange (DE) mechanism~\cite{zener,degennes}. However, the DE model
is incomplete to describe the entire phase diagram observed
experimentally. For instance, the coupling of the electrons
with the lattice may be crucial
to account for the insulating properties above
$T_c^{FM}$~\cite{millis}. The presence of a Berry phase 
in the large Hund coupling limit also challenges
 predictions 
from the DE model~\cite{muller}.
In this paper we remark that another phenomena occurring in manganites
which is
not included in the DE description, namely the charge ordering
effect, may be contained in a more fundamental Kondo
model where the $t_{2g}$ (localized) electrons 
are ferromagnetically (Hund) coupled with
the $e_g$ (mobile) electrons. More precisely, here we report the presence
of {\it phase separation} (PS) between hole undoped antiferromagnetic
and hole-rich ferromagnetic regions in the low
temperature phase diagram of the FM Kondo model. 
Upon the inclusion of long-range
Coulombic repulsion,  charge ordering
in the form of non-trivial extended structures 
may be stabilized. A similar phenomenon has been extensively discussed before
in the context of the high-Tc superconductors, although  without the presence of
ferromagnetic domains~\cite{tj1,review}. In the cuprates neutron scattering
experiments have shown indications of stripe 
formation~\cite{tranquada}, as predicted
by theoretical calculations after the inclusion of Coulomb 
interactions~\cite{tj2}. The analysis reported in this paper suggests
that phenomena as rich as  observed in the cuprates may exist in
the manganites, and hopefully the present
results will induce further theoretical and
experimental work in this context. 

The FM Kondo Hamiltonian~\cite{zener,furukawa} is defined as 
$$
H = -t \sum_{\langle {\bf i j} \rangle \sigma} (c^\dagger_{{\bf i}\sigma} c_{{\bf
j}\sigma} + h.c.) - J_H \sum_{{\bf i}\alpha \beta}
{ { c^{\dagger}_{{\bf i}\alpha} {\bf \sigma}_{\alpha \beta} c_{{\bf
i}\beta} }
\cdot{{\bf S}_{\bf i}}},
\eqno(1)
$$
\noindent where $c_{{\bf i}\sigma}$ are destruction
operators for one species of $e_g$-fermions at site ${\bf i}$
with spin $\sigma$,
and ${\bf S}_{\bf i}$ is the total 
spin of the $t_{2g}$ electrons, assumed 
localized.  
The first term is the $e_g$ electron transfer between nearest-neighbor
Mn-ions,
$J_H>0$ is the Hund coupling, the number of sites is $L$,
and the rest of the notation is standard. The density is adjusted using
a chemical potential $\mu$.
In this paper the spin ${\bf S}_{\bf i}$  will be considered classical
(with $|{\bf S}_{\bf i}| = 1$) rather than quantum mechanical,
unless otherwise stated. 
Phenomenologically $J_{H} \gg t$, 
but here $J_H/t$ was
considered an arbitrary parameter, i.e. both large and small values
for $J_H/t$ were studied.
Although models beyond Eq.(1) may be needed to fully understand
the manganites,
it is important to study the properties of simple
Hamiltonians  to clarify if part  of the experimental rich phase
diagram can be accounted for using
purely electronic models.

To study  Eq.(1) in the $t_{2g}$ spin classical
limit a Monte Carlo technique was
used: first, the trace over the fermionic
degrees of freedom in the partition function was carried out 
exactly diagonalizing the $2L \times 2L$ Hamiltonian of electrons
in the background of the spins \{${\bf S}_{\bf i}$\},
using library subroutines. 
Here the fermionic trace is a positive function of the classical spins
and, thus, the resulting integration over the two angles per site
parametrizing the ${\bf S}_{\bf i}$ variables can be performed with
a standard Monte Carlo algorithm without ``sign problems''.
In addition, part of the calculations were also performed with the
Dynamical Mean-Field approximation (D=$\infty$)~\cite{furukawa}, the
Density-Matrix Renormalization Group (DMRG), and the Lanczos method. 
Special care must be taken with
the boundary conditions (BC).
Closed shell BC or open BC are needed
to stabilize a ferromagnetic spin arrangement. If other BC are used the
spin correlations  at short distances
are still strongly FM (if working at couplings where ferromagnetism is
stable), but not at large distances where they become negative. This
well-known effect was observed before in this context~\cite{jose,zang}
and it does not present a problem in the analysis shown below.
\begin{figure}[htbp]
\centerline{\psfig{figure=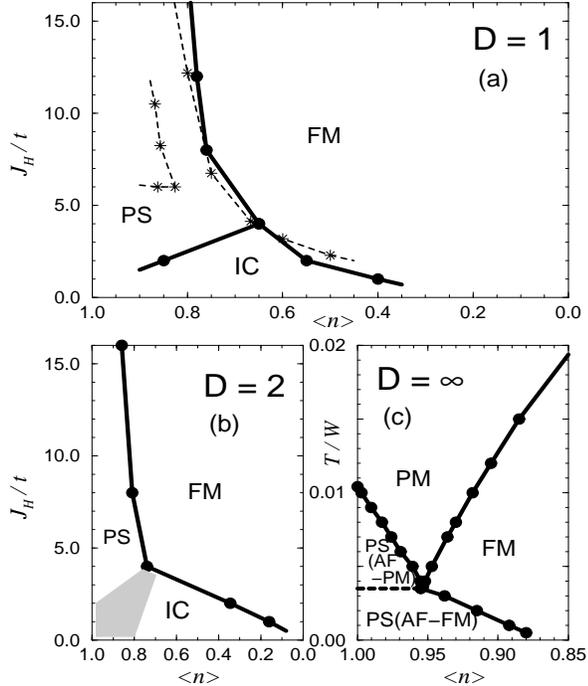,height=10cm,width=8.5cm}}
\vspace{0cm}
\caption{
Phase diagram of the FM Kondo model reported here.
FM, IC, PM and PS denote regimes with
strong FM correlations,  
incommensurate correlations,  paramagnetic correlations,
and with
phase separation between undoped AF
and hole-rich FM regions, respectively. 
(a) was obtained with MC simulations 
at  $T=t/75$ using chains with
$L$=20, 30 and 40 sites. The full
circles correspond to classical spins.
The stars and dashed
lines are DMRG results obtained with $t_{2g}$ spins
S=3/2 on chains with
up to $L$=16 sites keeping 48 states, and using $(2/3)J_{H}$ as
coupling in the Hund term;
 (b) are MC results for 
classical spins at $T=t/50$ using $4 \times 4$, $6 \times 6$ and
$8 \times 8$ clusters. In the shaded region a crossover from
PS to IC was observed but the actual boundary position is difficult
to find accurately; (c) corresponds to results
in the D=$\infty$ limit at a fixed coupling $J_H/W = 4.0$ and as
a function of $T$ and $\langle n \rangle$ ($W$ is defined in the text). 
The ``PS (AF-PM)'' region
denotes PS between undoped AF and hole-rich PM regions.
}
\vspace{0.5cm}
\end{figure}

The results reported here can be summarized in 
the  phase diagrams obtained for several dimensions D
shown in Fig.1. In both 1D and 2D and at low temperatures
clear indications of
(i) strong ferromagnetism, (ii) incommensurate (IC) correlations, and (iii)
a novel regime of phase separation were identified. 
For D=1 and 2, finite size effects were found to be small for the lattice
sizes used in this study, although the PS-IC boundary
in 2D was difficult to identify accurately and, thus, only a crossover
is indicated. Results are also available in small 3D clusters and
qualitatively they agree with those in Figs.1a-b. 
In 1D we also obtained results with quantum $t_{2g}$ spins
S=3/2.
In this case the PS regime was studied
calculating the compressibility with the DMRG technique, 
and the FM regime monitoring the
ground state spin quantum number with the Lanczos method. 
The agreement with
results in the classical limit is good, specially regarding the
ferromagnetic line. PS for S=3/2 is between
hole undoped AF and hole-rich non-fully-saturated FM regions.
Similar results were obtained in the case S=1/2.
In Fig.1c 
results in D=$\infty$ are shown at $J_H/W = 4$, where $W$ is
the half-width of the semicircular density of states 
$D( \epsilon ) = (2 / \pi W) \sqrt{ 1 - (\epsilon / W)^2}$
for the
 $e_g$ electrons. In excellent agreement with the
predictions for  D=1 and 2, at low temperatures phase
separation between undoped AF and hole-rich
FM regions was observed. For $\langle n
\rangle < 0.88$ and $T/W \ll 1$ the ground state becomes ferromagnetic.
The quantitative similarities between the results obtained for D=1,2 and
$\infty$ led us to believe that the conclusions of this paper
 are also valid for
3D manganite systems. 
Below the details justifying the proposed phase
diagrams shown in Fig.1 are provided.
\begin{figure}[htbp]
\vspace{-0.5cm}
\centerline{\psfig{figure=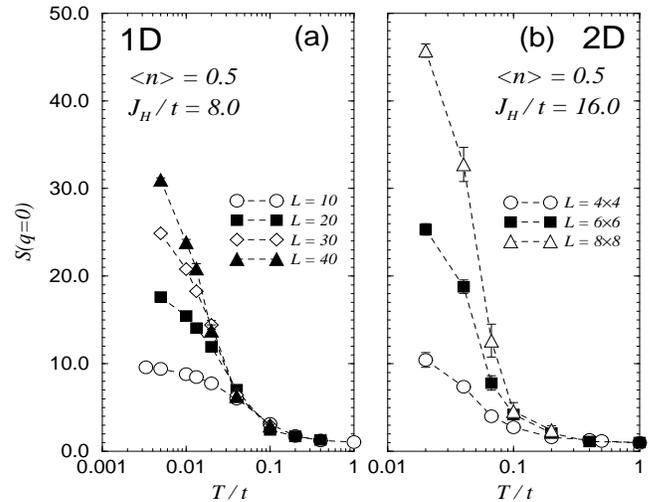,height=10cm,width=8.5cm}}
\vspace{-2.0cm}
\caption{
Spin-spin correlations of the classical spins
 at zero momentum $S({\bf q}=0)$
vs temperature $T$ (in units of $t$).
Results for several lattice
sizes are shown on (a) chains and (b) 2D clusters.
The technique used is the Monte Carlo algorithm described in the text.
The density and coupling are shown.
In (a) closed shell BC are used i.e. 
periodic BC for $L=10$ and $30$ and antiperiodic
BC for $L=20$ and $40$. In (b) open
BC are used.
}
\end{figure}

The boundaries of the FM region of the phase diagram 
were found evaluating the spin-spin correlation
between the classical spins defined as $S({\bf q}) = 
(1/L) \sum_{\bf j,m} e^{i {{\bf (j-m)}\cdot{\bf q}}}
\langle {{{\bf S}_{\bf j}}\cdot{{\bf S}_{\bf m}} } 
   \rangle$. Fig.2
shows $S({\bf q})$ at zero momentum 
vs. $T/t$ for typical examples in 1D and 2D.
The rapid increase of the spin correlations as $T$ is reduced and
as the lattice size grows clearly points towards the existence of
ferromagnetism in the system~\cite{foot3}. 
Repeating this procedure for a variety of couplings and densities,
the robust region of FM shown in Fig.1 was determined,
in qualitative
agreement with previous studies at $D=\infty$~\cite{furukawa} and
in the large $J_H/t$ limit on 1D chains~\cite{jose}.

In the small $J_H/t$ region incommensurate correlations were
observed. IC effects were predicted 
using a Hartree-Fock approximation~\cite{inoue}
as an alternative to canted
FM~\cite{degennes}.
The IC region was observed in our studies monitoring
$S({\bf q})$.
Both in 1D and 2D there is one
dominant peak which moves away from  the
AF location  at $\langle n \rangle =1$ towards zero momentum as $\langle
n \rangle$ decreases. In the 2D clusters
the peak moves from 
$(\pi,\pi)$ towards $(\pi,0)$ and $(0,\pi)$, rather than along
the main diagonal.  
Our results confirm a tendency towards the development of IC
correlations  in the FM Kondo model~\cite{inoue,hamada}.
However, our computational study predicts IC correlations only
in the small and intermediate 
$J_H/t$ regime i.e. in a region of parameter space
not realized in the manganites. Details will be provided elsewhere.
\begin{figure}[htbp]
\vspace{0cm}
\centerline{\psfig{figure=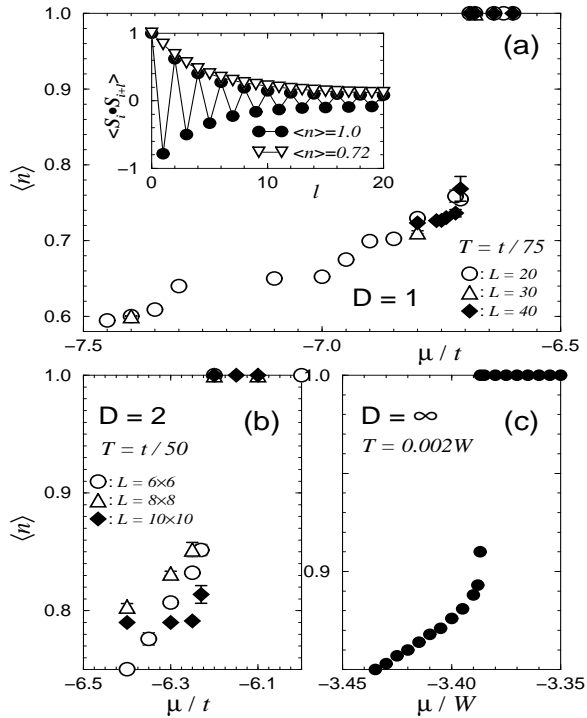,height=10cm,width=8.5cm}}
\vspace{0cm}
\caption{
Electronic density $\langle n \rangle$ vs chemical potential $\mu$ 
in (a) D=1, (b) D=2, and (c) ${\rm D=\infty}$ clusters. The
temperatures are indicated.
The coupling is $J_H/t=8.0$ in (a) and (b) and $J_H/W = 4.0$ in (c). 
The meaning of $W$ is explained in the text. PBC were used
both in D=1 and 2. The discontinuities shown in the figures are indicative
of PS. In (a) the inset contains the spin-spin correlation in real space
at densities 1.0 and 0.72 
showing that indeed PS occurs between AF and FM regions.
}
\end{figure}

The main result of the paper is contained in 
Fig.3 where the computational evidence for the existence of 
phase separation in dimensions 1, 2, and $\infty$ 
is given~\cite{foot7}. The presence of a
discontinuity in $\langle n \rangle$ vs $\mu$  shows that some
electronic densities can not be stabilized.
If the system is 
nominally prepared with such density it will spontaneously
separate into two regions with the densities corresponding to the 
extremes of the discontinuities of Fig.3~\cite{foot5}.
By analyzing these extremes
the properties of the two domains can be studied. 
At least in the 
classical limit for the $t_{2g}$ spins,
one region is undoped ($\langle n \rangle = 1$)
with strong AF correlations, while the other contains all the holes
and the spin-spin correlations between the classical spins
indicate the presence of strong FM correlations~\cite{foot2} (for a 
justification of this statement see the inset
of Fig.3a. The results are similar in D=2 and infinite).
This is natural since holes optimize their
kinetic energy in a fully aligned background. On the other hand, at 
$\langle n \rangle =1$ the DE mechanism is not operative: if the electrons
fully align their spins they simply cannot move in the conduction band 
due to the Pauli principle. Then, energetically it is better to form 
an antiferromagnetic pattern (similar intuitive and numerical
 conclusions were reached for the
case of S=3/2 and 1/2 localized spins).
As $J_H$ grows, the jump in Fig.3 is reduced and it tends to
disappear in the $J_H = \infty$ limit\cite{foot10}.

Experimentally, PS may be detectable using neutron diffraction
techniques if the two coexisting 
phases have different lattice parameters as in
${\rm La_2 Cu O_{4 + \delta}}$,
a ${\rm Cu}$-oxide 
with hole-rich and hole-poor regions~\cite{rada}. NMR and NQR
spectra, as well as magnetic susceptibility measurements,
can also be used to detect PS~\cite{hammel,bao}.
Note also that in the PS regime $S({\bf q})$ presents a
two peak structure, one located at the AF position and the other at zero
momentum. Since this also occurs in a canted ferromagnetic state care
must be taken in the analysis of the experimental data.
In particular recent experimental results by Kawano et al.~\cite{kawano} are in
qualitative agreement with Fig.1c since these authors observed a
reentrant structural phase transition accompanied by 
``canted ferromagnetism'' below $T_c^{FM}$, at $0.10 < x < 0.17$
in ${\rm La_{1-x} Sr_x Mn O_3}$.
Also the polaron-ordered
phase  reported by Yamada et al.~\cite{yamada} can be reanalyzed in terms of
the present results since it is known that the AF phase in 3D manganites
is orthorhombic while the FM
is pseudo-cubic. The formation of a lattice superstructure may stabilize the
magnetic tendency to phase separate and minimize lattice distortions.


Another alternative is that Coulombic forces and low $Sr$ mobility
prevent the macroscopic accumulation of charge intrinsic of a PS
regime. 
Thus, extended hole-rich domains in the form of stripes (as in cuprates)~\cite{tj2},
or some other arrangement, could be formed.
Although these details certainly
deserve further work, the results in this paper are enough to 
show that tendencies similar to those found
in models for the 2D cuprates, notably the $t-J$ model, may
also be operative in manganites.
The main difference is that instead of separation between
hole-poor AF and hole-rich superconducting regions, as in 
${\rm La_2 Cu O_{4 + \delta}}$, here the observed PS is between undoped AF
and hole-rich $ferromagnetic$ regimes. Tendencies to pairing in Cu-oxides
are replaced by tendencies to form ferromagnetic ground states in
Mn-oxides~\cite{jose}.

Although the phase diagrams of Fig.1 have PS 
close to half-filling, actually this phenomenon
should also occur at high hole doping $\langle n \rangle  \sim 0$
if an extra direct AF exchange interaction between the
localized $t_{2g}$ spins is
included. This coupling may be originated in a small hopping amplitude
for the $t_{2g}$ electrons. At $\langle
n \rangle =0$
model Eq.(1) supplemented by a Heisenberg  coupling $J'/t$
 among the localized spins
will certainly produce an AF phase, as in experiments,
 which upon electron doping will
induce a
competition between AF (with no $e_g$ electrons)
and FM electron-rich regions, similarly as in the half-filled
limit but replacing holes 
by electrons. Previous studies in 1D
support these claims~\cite{jose}. Thus, PS or charge ordering
could exist in manganites both at large and small fermionic densities.
\begin{figure}[htbp]
\vspace{-0.5cm}
\centerline{\psfig{figure=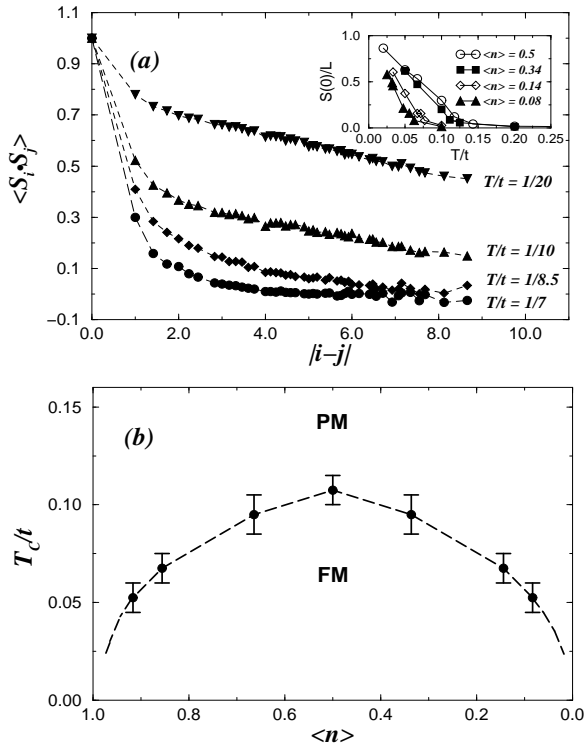,height=11cm,width=8.5cm}}
\vspace{0.5cm}
\caption{
(a) Real space spin-spin correlations among the classical spins
at distance $|i-j|$ obtained with
the MC technique at
$\langle n \rangle = 0.5$, $J_H = \infty$,
on a $6^3$ cluster, and parametric with temperature $T/t$. 
The inset shows $S({\bf q})$ at zero momentum and several densities vs $T/t$;
(b)
Bounds on the critical temperature as estimated from data as shown in
 (a) as a function of $\langle n \rangle$.
}
\end{figure}

For completeness, 
upper bounds on  the critical temperature $T_c^{FM}$ in 
3D systems are also provided. 
Using
Monte Carlo simulations in principle it is possible to calculate
$T_c^{FM}$  accurately. However, the characteristics of
the algorithm used here prevented us from 
studying clusters larger than $6^3$ even at $J_H=\infty$.
In spite of this limitation, 
monitoring the spin-spin correlations in real space 
allows us to judge at  what temperature  $T^*$
the correlation length reaches the boundary of
the $6^3$ cluster. Since the bulk $T_c^{FM}$ is smaller than $T^*$, this
allows us to establish upper bounds for the critical temperature based on
finite size cluster calculations. Fig.4a shows the spin-spin correlations
at several temperatures in the $J_H=\infty$ limit and at
$\langle n \rangle = 0.5$. 
When $T \sim 0.1t$ robust
correlations reach the boundary, while for $T \ge 0.12t$ the correlation
is short-ranged. Thus, at this density we estimate that
$T_c^{FM} < 0.12t$. 
Results for several densities in this
limit are shown in Fig.4b. 
For realistic densities, such as $\langle n
\rangle \sim 0.7$, our results are about a factor 1.7 smaller than
predicted by high temperature expansions~\cite{high} (this small
discrepancy may be caused by the use of S=1/2 localized spins 
in Ref.~\cite{high}). Nevertheless the order of magnitude of both 
calculations is similar.
Monitoring the rapid growth of the
 zero momentum spin correlations (inset of Fig.4a) provides similar bounds.
$T_c^{FM}$ is the highest at
$\langle n \rangle = 0.5$, if $J_H = \infty$. Since
results for the $e_g$ electrons bandwidth range
from $BW \sim 1~eV$~\cite{bandwidth} to $BW \sim 4 eV$~\cite{sarma},
producing a hopping $t = BW/12$ between $0.08$ and $0.33~eV$, then
our estimation for the critical temperature ranges roughly  between
$T_c^{FM} \sim 100~K$ and $400~K$.
This is within the range observed
experimentally,
in agreement with other 
results~\cite{furukawa,high}, and in disagreement with
previous estimations using classical spins
that predicted a much higher critical temperature 
$T_c^{FM} \sim 2000K$~\cite{millis}. Then, purely electronic models
can account for $T_c^{FM}$ without resorting to electron-phonon couplings.

Summarizing, here the main conclusion of the paper is that the
phenomenon of phase separation occurs in realistic models for
manganites. Then, experimentalists shoud consider this potential
regime in the interpretation of their results i.e. tendencies
to phase separate could be
as strong as those towards ferromagnetism in the real materials.
Hopefully the present paper will trigger discussion on
how phase separation will reveal itself in manganites from the
experimental point of view. It is anticipated that
the inclusion of
long-range Coulombic interactions may stabilize complex charge ordered
structures, that could be observed
experimentally using a variety of techniques.
Neutron scattering experiments should detect these charge fluctuations
above the critical temperature, similarly as it occurs in the
cuprates. From a more basic perspective,
our results suggest that the existence of phase separation 
in the context of electronic models for transition metal oxides
is a more general phenomenon than previously anticipated.

S. Y. is supported by the Japanese Society for the Promotion of Science.
J. H. is supported by the Florida State grant E\&G 502401002. He thanks
E. Miranda for useful discussions.
A. L. M. acknowledges the financial support
of the Conselho Nacional de Desenvolvimento Cient\'\i fico
e Tecnol\'ogico (CNPq-Brazil)
A. M. and E. D. are supported by the 
NSF grant DMR-9520776.

\medskip

\vfil

%
%


\begin{references}
\bibitem{foot3} In both 1D and 2D the Mermin-Wagner theorem forbids
a nonzero $T_c^{FM}$ in the model. However, the
correlation lengths can be very large even at finite $T/t$.
\bibitem{foot7} Up to $10^6$ sweeps were needed to obtain accurate
results in Fig.3.a-b, maintaining the acceptance rate close to 0.5.
\bibitem{foot5}  Actually, by fine tuning 
$\mu$ to be very close to its ``critical'' value, a two peak structure
was observed in $\langle n \rangle$ as in a first order phase transition.
\bibitem{foot2} It was observed that the 
FM correlations at the boundary with PS in Fig.1
are weaker than at the center of the FM phase, probably caused by
a small $T=0$ order parameter.
\bibitem{foot10} PS in manganites was also conjectured in Ref.\cite{jose}
based on 1D studies of the ``$t-J$'' version of Eq.(1).
\end{references}
\end{document}